\documentclass[preprint,showpacs,amsmath,amssymb,aps,floats,floatfix,nofootinbib]{revtex4-1}
\usepackage[colorlinks, linkcolor=blue, anchorcolor=red, citecolor=green]{hyperref}

\usepackage{bm} 
\usepackage{array}
\usepackage{graphicx}
\usepackage{subfigure}
\usepackage{ulem}

\newcommand\redsout{\bgroup\markoverwith{\textcolor{red}{\rule[0.5ex]{2pt}{0.4pt}}}\ULon}

\def\apj{Astrophys.\ J.\ }

\def\GeV{\,\mathrm{GeV}}
\def\MeV{\,\mathrm{MeV}}
\def\GV{\,\mathrm{GV}}

\def\cm{\,\mathrm{cm}}
\def\kpc{\,\mathrm{kpc}}
\def\fluxunitpern{\,\mathrm{m^{-2}s^{-1}sr^{-1} (GeV/n)^{-1}}}

\makeatother

\allowdisplaybreaks 

\begin{document}
\title{Expectations of the Cosmic Antideuteron Flux}
\author{Su-Jie Lin$^{1}$, Xiao-Jun Bi$^{1,2}$, Peng-Fei Yin$^{1}$}
\affiliation{$^{1}$Key Laboratory of Particle Astrophysics,
Institute of High Energy Physics, Chinese Academy of Sciences,
Beijing 100049, China  \\
$^{2}$School of Physics, University of Chinese Academy of Sciences, Beijing 100049, China}

\begin{abstract}
The cosmic antideuteron is a promising probe for the dark matter annihilation signature.
In order to determine the DM signature, the background astrophysical antideuteron flux should be carefully studied.
In this work we provide a new calculation of the secondary antideuteron flux, and pay special attention to the uncertainties from hadronic interaction models by using several Monte Carlo generators.
The uncertainties from propagation effects are also carefully investigated for both the astrophysical background and DM annihilation signature in several scenarios, which are constrained by the latest B/C ratio measured by AMS-02.
Considering these uncertainties, we find that the secondary antideuteron flux is hard to detect in the near future detectors.
However, the antideuteron signature from dark matter annihilation will be detectable even considering the constraint from the AMS-02 observation of the $\bar{p}/p$ ratio.
\end{abstract}
\maketitle

\section{Introduction}\label{section_Introduction}

The cosmic antideuteron can be generated by collisions between the high energy CR particles and interstellar gas~\cite{Fradkin:1955jr}.
However, the secondary antideuteron flux from this process is expected to be lower than the sensitivities of experiments~\cite{Duperray:2005si,Ibarra:2013qt,Herms:2016vop,Tomassetti:2017izg,Aramaki:2015pii,Dal:2015sha}.
On the other hand, dark matter (DM) annihilations or decays can also produce cosmic antideuterons in the Galaxy~\cite{Silk:1984zy}.
This exotic antideuteron flux may be much larger than the astrophysical background at low energies below $\sim 10$ GeV, and can be detected by future experiments~\cite{Donato:2008yx,Korsmeier:2017xzj,Grefe:2015jva,Dal:2014nda}.
Thus, the cosmic antideuteron is a promising probe for the indirect detection of DM\@.

No cosmic antideuteron has been observed up to now.
The best available limit is set by BESS\@, which is $\Phi_{\bar{d}} < 1.9\times10^{-4}\fluxunitpern$ in the range of $0.17\sim1.15\GeV/\mathrm{n}$ (kinetic energy per nucleon)~\cite{Fuke:2005it}.
The ongoing and future CR experiments will have much better sensitivities to the cosmic antideuteron flux.
After a five-year operation, the sensitivities of the Alpha Magnetic Spectrometer 02 (AMS-02) experiment are expected to reach $\Phi_{\bar{d}} = 2\times10^{-6}\fluxunitpern$ and $\Phi_{\bar{d}} = 1.4\times10^{-6}\fluxunitpern$ respectively within the energy ranges of $0.2\sim0.8\GeV/\mathrm{n}$ and $2.2\sim4.4\GeV/\mathrm{n}$~\cite{Aramaki:2015pii}, respectively.
Additionally, the balloon borne General Antiparticle Spectrometer (GAPS), which is planned to undertake a series of flight above Antarctica, will search for cosmic antideuterons in the energy window of $0.05\sim0.25\GeV/\mathrm{n}$ with a sensitivity of $\Phi_{\bar{d}}=2\times10^{-6}\fluxunitpern$~\cite{Aramaki:2015laa}.
Since the sensitivities of these experiments are much better than those of previous experiments, it is necessary to study the cosmic antideuterons flux potentially detected by them.

In order to search for the antideuteron signature from DM, the astrophysical background should be carefully predicted.
However, there remain many uncertainties arising from the hadronic production and CR propagation processes in the relevant calculation.
In this work, we take into account the impact of these uncertainties, and calculate the cosmic antideuteron fluxes for both the astrophysical background and DM annihilation signature.

The CR propagation process in the Galaxy propagation halo is described by a complicated diffusive function, which may involve a reacceleration and/or a convection effect~\cite{Moskalenko:2001ya}.
The degeneracy of these two effects leads to an important uncertainty in the propagation calculation in the low energy range.
In addition, the size of the Galaxy propagation halo is also not clearly determined due to the lack of precise measurements of unstable-to-stable secondary nuclei ratios.
In order to study the impact of propagation uncertainties, we consider four kinds of propagation models in the analysis, namely the diffusion-convection (DC) model, the diffusion-reacceleration (DR) model, the modified diffusion-reacceleration (DR-2) model, and the diffusion-reacceleration-convection (DRC) model.
We utilize the GALPROP package to solve the CR propagation equation~\cite{Strong:1998pw,Moskalenko:1997gh}.
The propagation parameters are determined by a Markov Chain Monte Carlo (MCMC,~\cite{Lewis:2002ah}) fitting to the latest AMS-02 B/C data.

Astrophysical background antideuterons are formed by antiprotons and antineutrons produced by the collisions between the high energy CR particles and the interstellar gas.
Therefore, the secondary production rates of cosmic antiprotons and antineutrons are important for the prediction of background antideuterons.
Since the forward hadronic scatterings can not be perturbatively calculated from the first principle, many parameterized methods~\cite{Tan:1984ha,Duperray:2003bd,diMauro:2014zea} and phenomenological Monte Carlo (MC) generators~\cite{Corcella:2000bw,Sjostrand:2007gs,Ostapchenko:2013pia,Gleisberg:2008ta,Kalmykov:1997te,Engel:1995yda,Pierog:2009zt,Pierog:2013ria} have been adopted to deal with these processes.
In this work, the production processes of cosmic antiprotons and antineutrons are simulated by the MC generators EPOS and QGSJET-II developed in the pomeron scenario.
These generators have been tuned to explain the available collider data and are widely adopted in CR studies.
For each event in the simulation, the formation of antideuteron is calculated by using the coalescence model.
Considering the uncertainties from the antideuteron production and propagation processes, we find that the secondary antideuteron flux is hard to detect in the near future detectors.

For the DM signature, we also adopt a MC method as discussed in Ref.~\cite{Kadastik:2009ts} by using the generator PYTHIA~\cite{Sjostrand:2007gs}.
The latest constraint from the AMS-02 antiproton observation on the DM annihilation cross section has been taken into account in the analysis.
We find that the antideuteron signature from dark matter annihilation will be detectable even considering the AMS-02 antiproton constraint.

This paper is organized as follows.
In Section~\ref{section_antideuteron_production}, we describe the estimation of the cosmic antideuteron production.
We also discuss the difference between the results given by different MC generators.
In Section~\ref{section_propagation}, we introduce the propagation models adopted in this work.
Then we calculate the antideuteron fluxes both for the astrophysical background and DM annihilation signatures in Section~\ref{section_results}.
Finally, we give our conclusion in Section~\ref{section_conclusion}.

\section{Antideuteron Production}\label{section_antideuteron_production}
In this work, we estimate the production of antideuterons via an MC simulation.
First, we generate antiprotons and antineutrons via MC generators for each collision event.
Then, we calculate the formation of antideuterons by using the coalescence model.

\subsection{Hadronic Interaction Models}\label{subsection_hadronic_interaction_models}
Although the Quantum Chromodynamics (QCD) has been well established and tested in many years, it could not yet directly predict the bulk of produced particles in hadronic interactions.
In order to deal with the relevant processes, many analytically parameterized methods~\cite{1983JPhG....9.1289T,1983PhRvC..28.2178L,1997APh.....6..379M} and MC generators have been developed in the literature~\cite{Corcella:2000bw,Sjostrand:2007gs,Ostapchenko:2004ss,Kachelriess:2015wpa,Engel:1992vf,Bopp:2005cr,Werner:2012xh,Pierog:2013ria}.
In this work, we adopt the MC generators QGSJET-II and EPOS to calculate hadronic interactions between the high energy primary CR particles and the interstellar gas here.
These generators are developed in the pomeron scenario, and have been tuned to fit the available collider data.

The QGSJET-II generator~\cite{Ostapchenko:2004ss,Ostapchenko:2005nj} is designed to simulate the extensive air showers induced by very high energy CR particles.
After reproducing the current collider data, a reasonable extrapolation method is used to deal with hadronic interactions in the high energy range and phase-space regions without available experiment results.
Although the predictions of antiprotons given by its early versions are not quit consistent with the NA49 results~\cite{Baatar:2012fua}, Kachelriess et al.~\cite{Kachelriess:2015wpa} have proposed a modified version to deal with this problem.
Thus we adopt this modified version of QGSJET-II as a typical choice in our analysis.

EPOS~\cite{Werner:2005jf,Pierog:2013ria} is another widely adopted hadronic generator in CR studies, which can reproduce the collider data well over a wide energy range.
Note that there are two popular versions of EPOS, namely EPOS 1.99~\cite{Werner:2005jf} and EPOS LHC~\cite{Pierog:2013ria}, which have been tuned to explain different collider data.
EPOS 1.99 focused on the low energy data, while EPOS LHC can well explain the recent high energy LHC results since 2009.
We adopt both versions of EPOS in our calculation.

\begin{figure}[!htp]
  \centering

  \includegraphics[width=0.5\textwidth]{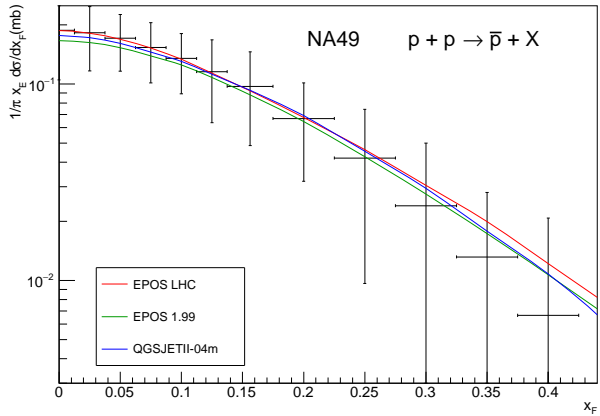}
\caption{ The antiproton differential cross section $1/\pi x_E \mathrm{d}\sigma/\mathrm{d}x_F$~.vs.
$x_F$ estimated by the MC generators QGSJETII-04m, EPOS 1.99 and EPOS LHC, comparing with the NA49 data at beam momentum $158\GeV/c$~\cite{Baatar:2012fua}.
Both the $x_F\equiv 2 p_z/\sqrt{s}$ and $x_E\equiv 2E/\sqrt{s}$ are defined in the CM system.
}
\label{fig:pbar_cs}
\end{figure}

Both the EPOS and QGSJET generators can provide a reasonable prediction for hadronic collisions with center-of-mass (CM) energies $\sqrt{s}$ from tens of $\GeV$ to hundreds of $\GeV$.
As can be seen in Fig.~\ref{fig:pbar_cs}, their predictions for the antiproton production well fit the NA49 data with a beam momentum of $158\GeV/c$.
For the antiproton production at low energies, there is a set of results from the S61 experiment with an incident proton momentum of $19.2\GeV$ (equivalent to a $\sqrt{s}=17.3\GeV$) at CERN PS~\cite{Allaby:1970jt}.
The corresponding CM energy $\sqrt{s}=6.15\GeV$ is close to the antideutron production threshold $\sim 6 m_p$.
In Ref.~\cite{Kachelriess:2015wpa}, the predicted antiproton spectra given by QGSJET-IIm and EPOS LHC have been compared with the S61 data.
The authors found that QGSJET-IIm can well explain the measured antiproton spectra below $\sim 8$GeV, while the spectra predicted by EPOS are slightly lower in the same energy region.
Although the capabilities of these generators at low CM energies are not as well as those at high CM energies, their predictions are acceptable to study the low energy antiproton and antineutron production in secondary CRs.

\begin{figure}[!htp]
  \centering
  \includegraphics[width=0.5\textwidth]{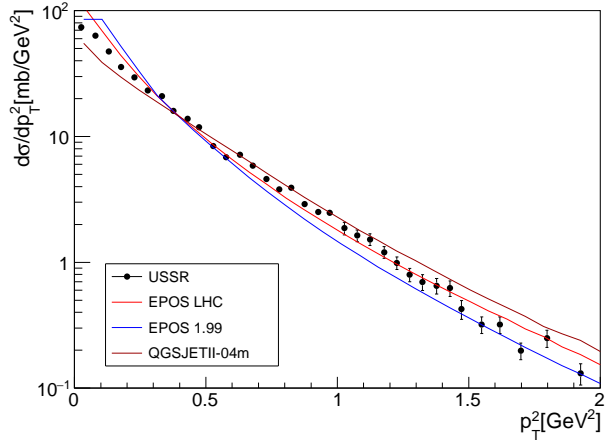}
\caption{ The $\bar{p}+p\rightarrow p + X$ cross section at beam momentum $32\GeV/c$.
The lines show the prediction for different generators (after rescaled), and the data points are the measurements at the Mirabelle bubble chamber at IHEP~\cite{Vlasov:1982rm}.
}
  \label{fig:ussr}
\end{figure}

For proton-antiproton collisions, we compare the $\bar{p} + p\rightarrow p + X$ cross section predicted by the MC generators with the results of Mirabelle bubble chamber at IHEP~\cite{Vlasov:1982rm} with a beam momentum of $32\GeV/c$.
As shown in Fig.~\ref{fig:ussr}, we find that the generators well reproduce the experimental data after introducing some rescale factors.
Thus we apply these rescale factors for the low energy proton-antiproton interactions in the following calculation.

\subsection{Coalescence Model}\label{subsection_coalescence_model}

All the MC generators mentioned above are only used to simulate the production of antiprotons and antineutrons.
In order to simulate the production of antideuterons, we adopt the so called coalescence model~\cite{Csernai:1986qf}, in which an antideuteron is formed by a pair of antiproton and antineutron when their momentum difference $\Delta p$ is smaller than the coalescence momentum $p_0$.

\begin{figure}[!htp]
  \centering
  \includegraphics[width=0.5\textwidth]{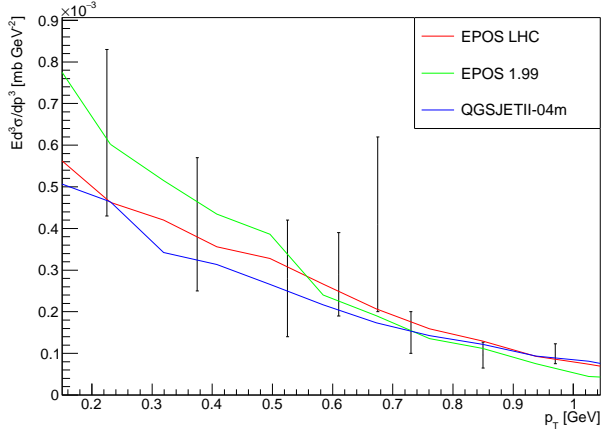}
\caption{The differential cross sections of antideuterons comparing with the data measured at CERN ISR~\cite{Alper:1973my,Henning:1977mt} for the chosen generators with best-fit $p_0$ values.
$p_T$ is the transverse momentum of antideuteron.}
\label{fig:CERN_ISR}
\end{figure}

The $p_0$ is a phenomenological parameter and can be determined by fitting experimental data.
In principle, its value varies not only for different generators, but also for different collision energies.
Here we just assume that the value of $p_0$ does not vary with the collision energy for simplicity.
We use the differential cross section of antideuterons measured by CERN ISR~\cite{Alper:1973my,Henning:1977mt} with $\sqrt{s}=53\GeV$ to determine $p_0$.
The best-fit values of $p_0$ are given in Table.~\ref{tab:p_0}.
The corresponding differential antideuteron production cross sections predicted by different generators are shown in Fig.~\ref{fig:CERN_ISR}.

\begin{table}
  \centering
  \begin{tabular}{ccccc}
    \hline
    \hline
                 & EPOS LHC & EPOS 1.99 & QGSJETII-04m & PYTHIA 8 \\
    \hline
     $p_0(\MeV)$ & 175.8 & 145.3 & 211.7 & 192 \\
    \hline
    \hline
  \end{tabular}
\caption{The $p_0$ values.
The values for EPOS and QGSJET are obtained by fitting with the data at CERN ISR~\cite{Alper:1973my,Henning:1977mt}, while the $p_0$ value for PYTHIA 8 is adopted from Ref.~\cite{Ibarra:2012cc}.}
  \label{tab:p_0}
\end{table}

\subsection{Injection of Background Antideuteron}\label{subsection_injection_of_cr_antideuteron}
With the derived antideuteron production rate, we perform an estimation for the injection of the background antideuteron induced by interactions between the CR particles and interstellar gas.
Its injection can be presented by
\begin{equation}
  Q_j(p)=\beta c n_{gas}\int \mathrm{d}p'\frac{\mathrm{d}\sigma_{i\rightarrow j}(p'\rightarrow p)}{\mathrm{d}p}n_i(p'),
  \label{secondary_injection}
\end{equation}
where the index $i$ (or $j$) indicates the specie of particles, $\sigma_{i\rightarrow j}$ is the cross section of the corresponding production process, $\beta$ is the velocity of the secondary particle in unit of $c$, and $n_{\mathrm{gas}}$ and $n_{i}$ are the densities of the interstellar gas and the $i$-th CR particles, respectively.

There are two processes that would contribute to the background antideuteron injection.
The primary CR particles result in the secondary antideuterons and antiprotons.
These secondary antiprotons can interact with the interstellar gas again and produce the tertiary antideuterons.
In the estimation of these two processes, we adopt the local interstellar (LIS) fluxes of the primary proton and secondary antiproton predicted by the DR model in Ref.~\cite{Lin:2014vja}; the interstellar hydrogen density is taken to be $n_H=1.0\mathrm{cm}^{-3}$.

\begin{figure}[!htp]
  \centering
  \includegraphics[width=0.5\textwidth]{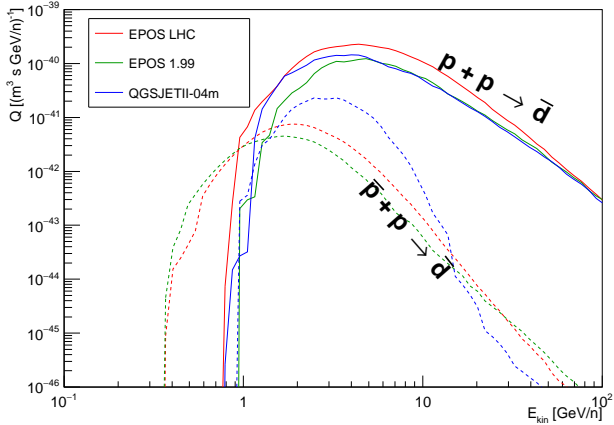}
\caption{The injection of CR secondary antideuteron around the local interstellar.
The solid line indicate the secondary antideuteron (from proton-proton collision) while the dash line indicate the tertiary antideuteron (from proton-antiproton collision)}
  \label{fig:injection}
\end{figure}

The injection spectra for different generators are shown in Fig.~\ref{fig:injection}.
When the kinetic energy is around or lower than $\sim \GeV$, the contribution of tertiary antideuteron may be dominant for the final injection.
Although the secondary antiprotons are rare compared with the primary protons, their interactions with the interstellar gas around the antideuteron production threshold are much stronger than those of the primary protons.
In addition, there are low energy cutoffs in both the secondary and tertiary injection spectra.
This is because that the secondary and tertiary antideuterons tend to be produced with low energies in the CM frame.
Thus their energies in the realistic frame would always be larger than a minimum value depending on the threshold of interaction.

Note that except for the injection induced by the primary protons, the contributions from CR heliums, whose abundance is about $10\%$ of the primary proton abundance, should also be taken into account.
Instead of directly simulating the antiproton-helium, proton-helium and helium-helium interactions with the generators again, we derive them from the proton-antiproton and proton-proton interactions for simplicity.
We perform this calculation with the method provided by Ref.~\cite{1992ApJ...394..174G}.
The collision cross sections of the nuclei (antinuclei) with nucleon numbers of A and B can be derived by
\begin{equation}
  \begin{split}
    \sigma_{AB\rightarrow\bar{d}} &\simeq \frac{1}{2}w_{AB}\frac{\sigma_{AB,\mathrm{inel}}}{\sigma_{pp,\mathrm{inel}}}\sigma_{pp\rightarrow\bar{d}} \\
                                  &= \frac{1}{2}\left( B\frac{\sigma_{pA,\mathrm{inel}}}{\sigma_{pp,\mathrm{inel}}} + A\frac{\sigma_{pB,\mathrm{inel}}}{\sigma_{pp,\mathrm{inel}}} \right) \sigma_{pp\rightarrow\bar{d}},
  \end{split}
  \label{effective_nuclear_factors}
\end{equation}
where the $\sigma_{*,\mathrm{inel}}$ is the total inelastic cross section for the corresponding collision, and the $w_{AB}$ is a raw estimation on the total number of nuclei that have effects on the collisions.
In our analysis, these total inelastic cross sections are evaluated by the CROSEC code of Barashenkov-Polanski embedded in the GALPROP package~\cite{Strong:1998pw}.

\section{Propagation}\label{section_propagation}
After being injected to the Galaxy, the antideuterons travel in the propagation halo, which is a cylindrical area around the Galaxy disk with a thickness of several $\kpc$.
The propagation effects are described by a diffusive transport equation expressed by
\begin{equation}
  \begin{split}
    \frac{\partial \psi}{\partial t} =& Q(\mathbf{x}, p) + \nabla \cdot \left( D_{xx}\nabla\psi - \mathbf{V}_{c}\psi \right)
    + \frac{\partial}{\partial p}p^2D_{pp}\frac{\partial}{\partial p}\frac{1}{p^2}\psi \\
    &- \frac{\partial}{\partial p}\left[ \dot{p}\psi - \frac{p}{3}\left( \nabla\cdot\mathbf{V}_c\psi \right) \right]
    - \frac{\psi}{\tau_f} - \frac{\psi}{\tau_r},
  \end{split}
  \label{propagation_equation}
\end{equation}
where the $Q$ is the source term.
The terms with $D_{xx}$, $\mathbf{V}_c$, $D_{pp}$, $\dot{p}$, $\tau_f$, and $\tau_r$ describe the diffusion, convection, diffusive reacceleration, energy loss, fragmentation, and radioactive decay effects, respectively.
We solve this equation with the numerical package GALPROP~\cite{Strong:1998pw}.

The diffusion coefficient $D_{xx}$ is assumed to vary with the rigidity $R$ of the CR particle, following the relation of $D_{xx}=D_0\beta^\eta(R/R_0)^\delta$, where $\beta$ is the velocity of the particle in unit of $c$, and $R_0=4\GV$ is the reference rigidity.
The constant $\eta$ is taken to be 1 by default, but it was found that a negative value of $\eta$ may describe the low energy behaviour of $D_{xx}$ better~\cite{DiBernardo:2010is}.

The propagation parameters can be determined by the secondary-to-primary CR nuclei ratios like B/C and (Sc+Ti+V)/Fe together with the unstable-to-stable secondary ratios like $^{10}\mathrm{Be}/^9\mathrm{Be}$ and $^{26}\mathrm{Al}/^{27}\mathrm{Al}$~\cite{Strong:1998pw,1990cup..book.....G,DiBernardo:2009ku,Maurin:2001sj}.
However, as the current measurements of unstable-to-stable secondary ratios are not precise enough, the determination of propagation parameters dominantly depend on the secondary-to-primary ratios, especially, the B/C ratio.
These ratios are just related to the average propagation distance of CR particles in the Galaxy, which depends on both the $D_{xx}$ and the half-thickness of propagation halo $L$.
Therefore, the values of $L$ and $D_{xx}$ determined by this approach are always degenerate.
The value of $L$ has been found within a large range of $2\sim15\kpc$ in the fitting to the current B/C data~\cite{Yuan:2015rlv,Yuan:2017ozr}.

The convection effect is induced by the galactic wind of charged particles, and can be described by a convective velocity $\mathbf{V}_c$ of the background environment.
In this work, the convection velocity is assumed to be in proportion to the distance from the Galaxy disk $\mathbf{V}_c=z\cdot\mathrm{d}v/\mathrm{d}z$.
Another important propagation effect is the diffusive reacceleration effect caused by the random magenetohydrodynamic waves, which can change the momenta of charged particles.
This effect can be considered as a diffuse effect in the momentum space; the corresponding diffusion coefficient is given by~\cite{1990acr..book.....B,1994ApJ...431..705S}
\begin{equation}
  D_{pp}D_{xx}=\frac{4p^2v_A^2}{3\delta(4-\delta^2)(4-\delta)},
  \label{equation_Dpp}
\end{equation}
with $v_A$ is the Alfven velocity.
Note that the convection and reacceleration effects may not be simultaneously significant.
With either of them, the solutions of eq.~(\ref{propagation_equation}) are able to accommodate the current CR observations~\cite{Yuan:2017ozr}.
In this work, we consider four propagation models: the model only including the diffusion convection effect together with a $\delta=0$ below the reference rigidity $R_0$ (DC), the model only including the diffusion reacceleration effect with a $\eta=1$ (DR) or a free $\eta$ (DR2), and the model including both the convection and reacceleration effects (DRC).

For these propagation models, we adopt the propagation parameters from the most recent fitting results of Ref.~\cite{Yuan:2017ozr}.
This study took into account the latest AMS-02 $\mathrm{B/C}$ ratio and proton observations in the fitting at the same time, and calculated the corresponding nuclei injections following a broken power law as
\begin{equation}
    Q_{\mathrm{primary}}(R)\propto\left\{
    \begin{array}{ll}
      \left( R / R_{\mathrm{br}}^1\right)^{-\gamma_0}, & R < R_{\mathrm{br}} \\
      \left( R / R_{\mathrm{br}}^1 \right)^{-\gamma_1}, & R_{\mathrm{br}} < R \\
    \end{array}\right..
  \label{injection_equation}
\end{equation}
All the propagation parameters described above are listed in Table.~\ref{tab:parameters}.
In order to take into account the impact of the degeneracy between the diffusion coefficient and propagation halo height, we take DR2 as a typical model, and consider several cases with different values of the propagation halo height $L$ varying from $3\kpc$ to $9\kpc$.
These cases are the best-fit results for specific $L$s.
Among the four DR2 cases given in Table.~\ref{tab:parameters}, the global best-fit result is provided by the DR2-2 case~\cite{Yuan:2017ozr}.
\begin{table*}
  \centering
  \begin{tabular}{l*{14}{>{$}c<{$}}}
    \hline
    \hline
    & D_0 & \eta & L & v_A & \delta & R_0 & \mathrm{d}v/\mathrm{d}z & & A_{p}\footnote{Normalization at $100\GeV$ in unit of $10^{-9}\mathrm{cm}^{-2}\mathrm{sr}^{-1}\MeV^{-1}$} & \gamma_0 & \gamma_1 & R_{\mathrm{br}} & \Phi_+ \\
    \cline{2-8}\cline{10-14}
    & [10^{28}\cm^2/\mathrm{s}] &  & [\kpc] & [\mathrm{km/s}] &  & [\GV] & [\mathrm{km/s}\cdot\kpc] & &  &  &  & [\GV] & [\GV] \\
    \hline
DR2-1 & 2.48 & -2.05 & 3.0 & 15.6 & 0.524 & 4.0 & \text{---} & & 4.67 & 2.11 & 2.31 & 13.1 & 0.558 \\
DR2-2 & 4.16 & -1.28 & 5.02 & 18.4 & 0.5 & 4.0 & \text{---} & & 4.63 & 2.04 & 2.33 & 10.7 & 0.564 \\
DR2-3 & 5.26 & -1.43 & 7.0 & 17.9 & 0.5 & 4.0 & \text{---} & & 4.6 & 2.06 & 2.33 & 10.9 & 0.56 \\
DR2-4 & 6.38 & -1.0 & 9.0 & 19.2 & 0.485 & 4.0 & \text{---} & & 4.57 & 2.03 & 2.34 & 10.7 & 0.578 \\
\hline
DR & 7.24 & 1.0 & 5.93 & 38.5 & 0.38 & 4.0 & \text{---} & & 4.5 & 1.69 & 2.37 & 12.88 & 0.455 \\
DC & 4.95 & 1.0 & 10.8 & \text{---} & 0/0.591\footnote{Below/Above $R_0$} & 5.29 & 5.02 & & 4.61 & 2.43 & 2.3 & 60.26 & 0.686 \\
DRC & 6.14 & 1.0 & 12.7 & 43.2 & 0.478 & 4.0 & 11.99 & & 4.52 & 1.82 & 2.37 & 16.6 & 0.492 \\
\hline\hline
  \end{tabular}
  \caption{The propagation parameters, nuclei injection parameters, and the solar modulation potential for AMS-02 proton
    adopted from \cite{Yuan:2017ozr}.}
  \label{tab:parameters}
\end{table*}

These models can also reasonably explain the CR antiproton observations.
Note that the determination of the propagation parameters depend on the heavy-ion fragmentation cross sections~\cite{READ1984359} adopted in the fitting to secondary-to-primary ratios, especially the cross section of $\mathrm{C,O\rightarrow B}$.
Therefore, the derived propagation parameters are affected by the uncertainties from the measurements of heavy-ion fragmentation cross sections.
If such uncertainties are included, the derived propagation parameters and the predicted CR fluxes would be slightly modified.
Here we simply introduce a normalization factor $C_\mathrm{cs}$ for the proton-proton and antiproton-proton cross sections to compensate such an uncertainty.
Thus the secondary and tertiary fluxes would be scaled by a factor of $C_\mathrm{cs}$ and $C_\mathrm{cs}^2$, respectively.

We adopt the force field approximation (FFD)~\cite{Gleeson:1968zza}, which can be characterized by a solar modulation potential $\Phi$, to deal with the modulation effect for CRs propagation in the Solar system.
In order to consider the effects induced by the electric charge of CR particles~\cite{Maccione:2012cu,Potgieter:2013cwj}, we take two solar modulation potential $\Phi_-$ and $\Phi_+$ for the CR antiproton (antideuteron) and CR proton, respectively.

\begin{figure}[!htp] \centering
  \includegraphics[height=0.5\textwidth]{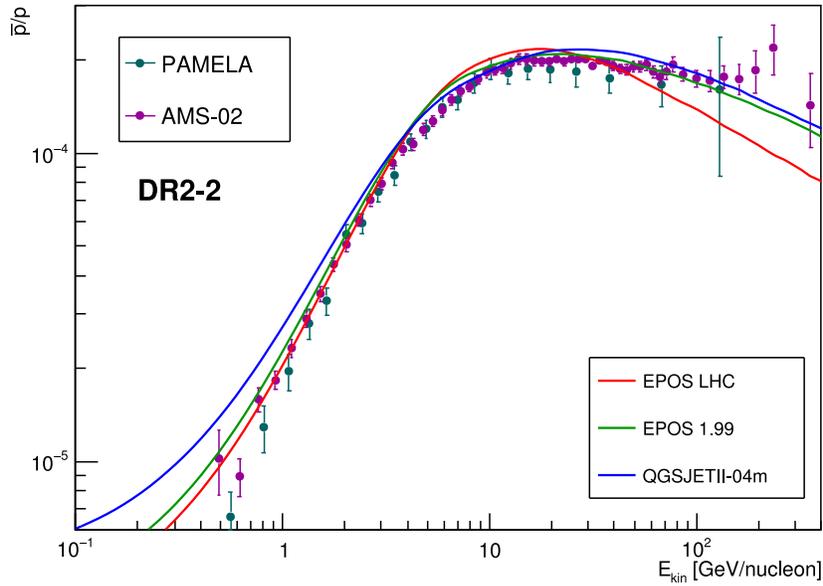}
\caption{The antiproton-to-proton ratio expected in DR2-2 case comparing with the AMS-02 \cite{Aguilar:2016kjl} and PAMELA \cite{Wu:2011zzs} results for different generators.
All the lines are the best-fit cases by varying the $C_\mathrm{cs}$ and $\Phi_-$, read the text for detail.}
\label{fig:pbarp}
\end{figure}

Then we can calculate the CR antiproton-to-proton ratio $\bar{p}/p$, and compare it with the observations given by AMS-02~\cite{Aguilar:2016kjl}.
Since only the low energy antiprotons are important for background antideuteron searches, we only focus on the antiproton-to-proton data below $50\GeV$.
The best-fit values of $C_\mathrm{cs}$ and $\Phi_-$ for different generators and propagation models are listed in Table.~\ref{tab:c_phi}.
We find that EPOS 1.99 and EPOS LHC can provide a better prediction of $\bar{p}/p$ than that given by QGSJETII-04m.
For instance, we show the predicted $\bar{p}/p$ for the DR2-2 case in Fig.~\ref{fig:pbarp}.
It seems that QGSJETII-04m overestimates the antiproton production below $\sim 10$ GeV and tends to require a large modulation potential $\Phi_-\sim$1.5 GeV for the DR2 model.
As there is no precise measurement on the cosmic antideuteron yet, such a slight overestimation are acceptable in this work.

\begin{table*}
  \centering
  \begin{tabular}{l*{3}{c>{$}c<{$}>{$}c<{$}}}
    \hline
    &  & \multicolumn{2}{c}{EPOS LHC} &  & \multicolumn{2}{c}{EPOS 1.99} & & \multicolumn{2}{c}{QGSJETII-04m} \\
    \cline{3-4}\cline{6-7}\cline{9-10}
    &  & C_\mathrm{cs} & \Phi_-[\GV] &   &  C_\mathrm{cs} & \Phi_-[\GV] &   & C_\mathrm{cs} & \Phi_-[\GV] \\
    \hline
DR2-1 & & 1.06 & 1.15 &  & 1.06 & 1.01 &  & 1.74 & 1.71\\
DR2-2 & & 1.02 & 0.956 &  & 1.03 & 0.827 &  & 1.68 & 1.58\\
DR2-3 & & 0.964 & 0.901 &  & 0.97 & 0.772 &  & 1.59 & 1.53\\
DR2-4 & & 0.963 & 0.821 &  & 0.971 & 0.701 &  & 1.6 & 1.47\\
\hline
DR & & 0.866 & 0.104 &  & 0.89 & 0.0455 &  & 1.47 & 0.952\\
DC & & 1.04 & 1.04 &  & 1.03 & 0.9 &  & 1.67 & 1.55\\
DRC & & 0.908 & 0.246 &  & 0.924 & 0.163 &  & 1.53 & 1.01\\
\hline\hline
  \end{tabular}
  \caption{The value of cross section scaling factor $C_\mathrm{cs}$ and modulation potential for AMS-02 antiproton $\Phi_-$}
  \label{tab:c_phi}
\end{table*}

\section{Results}\label{section_results}
In this section, we calculate the background antideuteron flux, and show the impact of the uncertainties from the production and propagation processes.
Then we discuss the antideuteron signatures from DM annihilations.

\subsection{Background Antideuterons}
\begin{figure}[!htp]
  \centering
  \includegraphics[height=0.5\textwidth]{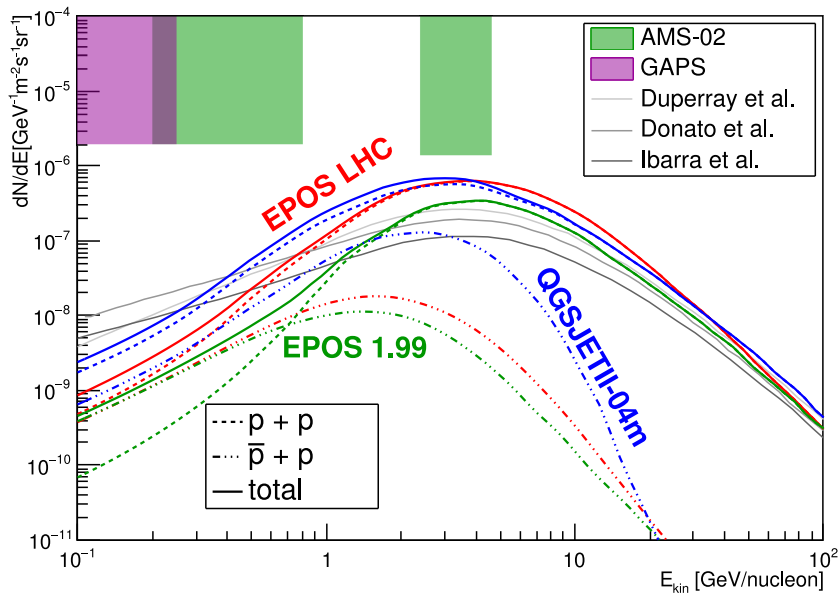}
\caption{The background antideuteron fluxes for different generators in DR2-2 case, comparing with the prospected sensitivity of AMS-02 and GAPS ULDB\@.
The dash lines, dot-dash lines and solid lines indicate the secondary, tertiary and total fluxes correspondingly.
The three gray lines are the result from Duperray et al.~\cite{Duperray:2005si}, Donato et al.~\cite{Donato:2008yx} and Ibarra et al.~\cite{Ibarra:2013qt}.}
\label{fig:dbar_bkg}
\end{figure}

First, we consider the impact of the hadronic interaction model on the background antideuteron flux, and show the results evaluated in the DR2-2 case in Fig.~\ref{fig:dbar_bkg}.
The different generators provide similar predictions for background antideuterons with energies larger than $\sim 10$ GeV, while their predictions are different at the low energy region.
In particular, the antideuteron flux evaluated from QGSJETII-04m is larger than that from EPOS 1.99 by a factor of $\sim 5$ in sub-GeV region and by a factor of $\sim 2$ around the peak of several GeV.

For a comparison, the results from Refs.~\cite{Duperray:2005si,Donato:2008yx,Ibarra:2013qt} are also shown as gray lines in Fig.~\ref{fig:dbar_bkg}.
The authors of Ref.~\cite{Duperray:2005si} and Ref.~\cite{Donato:2008yx} directly calculated the antideuteron production rate in the momentum space by introducing a phase space suppression factor, while the authors of Ref.~\cite{Ibarra:2013qt} performed a MC simulation similar to ours but with the generator DPMJET-III\@.
In Ref.~\cite{Ibarra:2013qt}, the authors found DPMJET-III overestimates the differential antiproton production cross section in the fitting to the S61 data.
They introduced a scaling factor as a function of the incident proton energy for the predicted antiproton production from DPMJET-III~\cite{Roesler:2000he}.
This factor is determined by comparing the DPMJET-III predictions to a interpolating function of experimental values~\cite{Antinucci:1972ib}, and is smaller than 1 for the incident proton energies below $\sim 400$ GeV.
We do not introduce such a scaling factor in our analysis, since the generator predictions adopted here are in consistent with the differential antiproton cross sections measured by S61 with $T_p=18.3\GeV$, NA49 with $T_p=158\GeV$, and CERN ISR with $T_p=1496.4\GeV$ ( a comprehensive discussion can be found in Ref.~\cite{Kachelriess:2015wpa}).
This is the reason why the predicted background antideuteron fluxes in this work are different from that of Ref.~\cite{Ibarra:2013qt} for the antideuteron kinetic energies of $\sim \mathcal{O}(1)$ GeV.

\begin{figure}[!htp]
  \centering
  \includegraphics[height=0.5\textwidth]{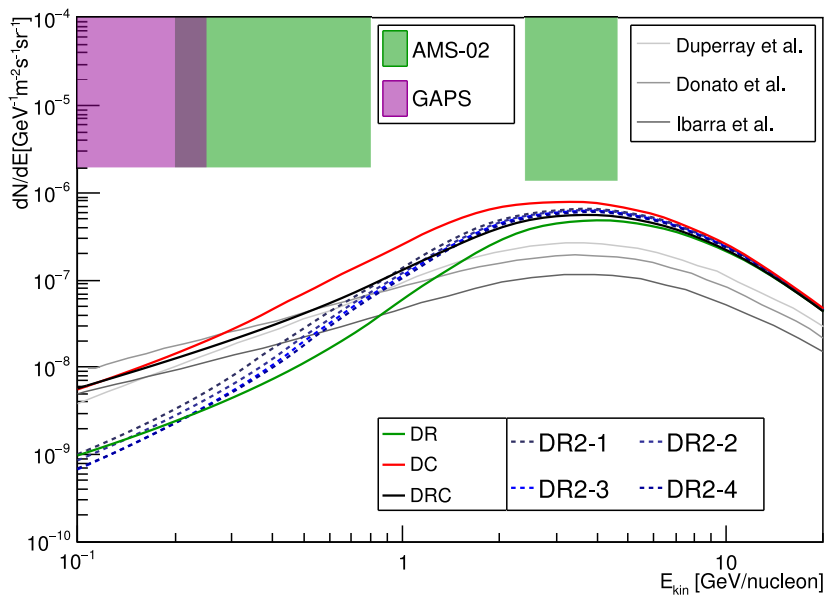}
\caption{The background antideuteron fluxes for different propagation parameters choices with generator EPOS LHC, comparing with the prospected sensitivity of AMS-02 and GAPS ULDB.
The colored lines indicate different propagation choices, while the three gray lines are the result from Duperray et al.~\cite{Duperray:2005si}, Donato et al.~\cite{Donato:2008yx} and Ibarra et al.~\cite{Ibarra:2013qt}.}
\label{fig:dbar_bkg_prop}
\end{figure}

Second, we consider the impact of the propagation model on the background antideuteron flux.
Taking the injection generated with EPOS LHC as a typical example, we show the results evaluated in different propagation models in Fig.~\ref{fig:dbar_bkg_prop}.
As can be seen that all results in the DR scenario coincide with each other.
This means that the variations of the propagation coefficient $D_0$ and halo size $L$ do not significantly affect the background antideuteron flux.
On the other hand, the different choices of the convection and reacceleration effects would lead to different results varying by a factor of $\sim 3-10$ at kinetic energies $E_\mathrm{kin}$ bellow 1 $\GeV$.

\begin{figure}[!htp]
  \centering
  \includegraphics[height=0.35\textwidth]{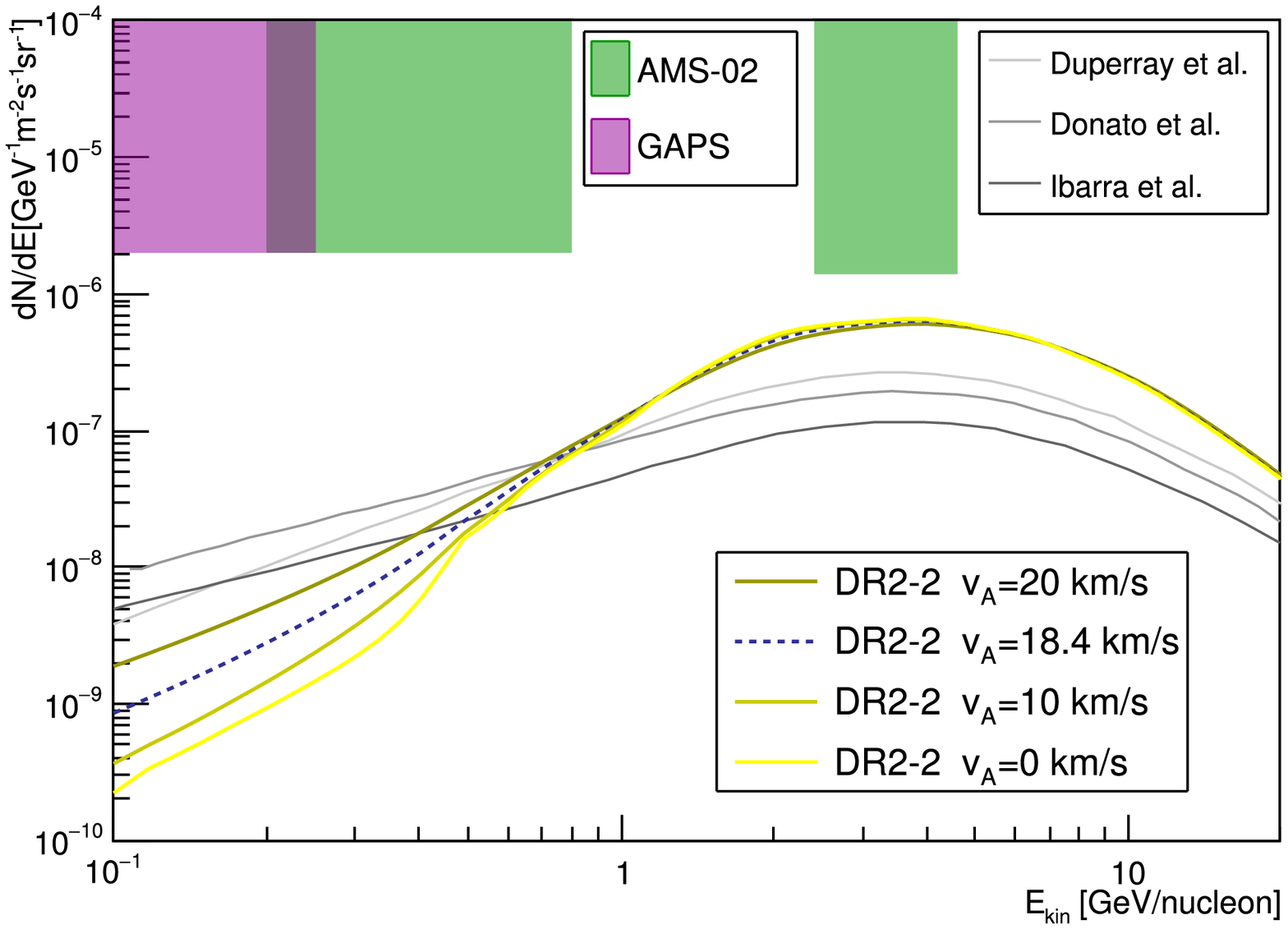}
  \includegraphics[height=0.35\textwidth]{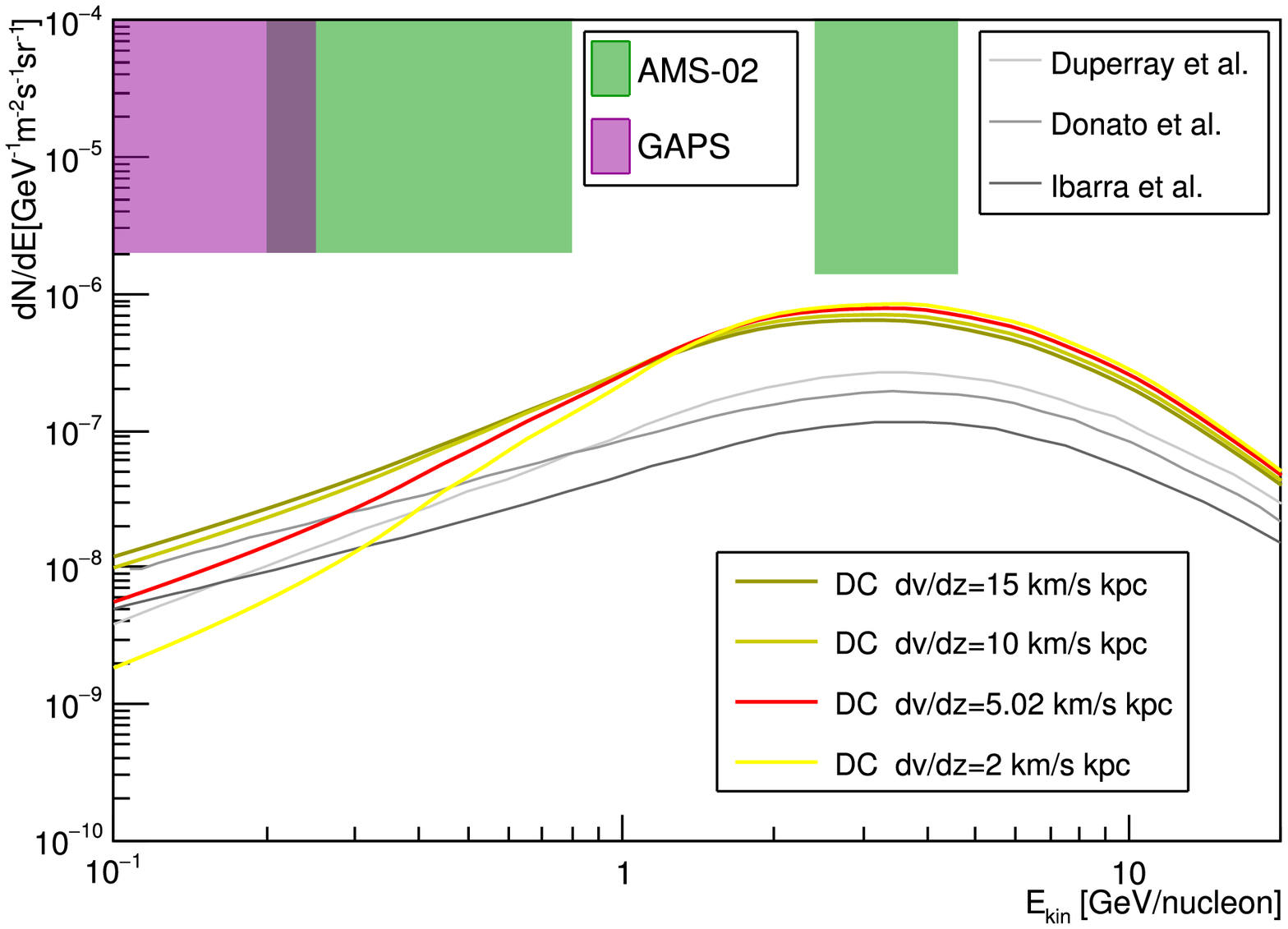}
\caption{The background antideuteron fluxes evaluated with different propagation parameters. The hadronic interaction generator is taken to be EPOS LHC.
In the left panel, the blue line is predicted by the typical DR2-2 case, while the yellow lines are predicted with different values of the Alfven velocity $v_A$.
The right panel is similar but for DC case and the gradient of convection velocity $\mathrm{d}v/\mathrm{d}z$.
}
\label{fig:dbar_bkg_prop_DRDC2}
\end{figure}

In Fig.~\ref{fig:dbar_bkg_prop_DRDC2}, we show the variation of the background antideuteron prediction for different $\mathrm{d}v/\mathrm{d}z$ and $v_A$.
Since the antideuteron injection below $\sim 1$ GeV is suppressed as shown in Fig.~\ref{fig:injection}, the observed antideuterons in the sub $\GeV$ range are dominantly injected at higher energies and lose their energies through the convection and reacceleration effects during the propagation.
Consequently, we can see that both the convection and reacceleration effects enhance the antideuteron flux at sub $\GeV$.

Adopting the different approaches for the convection and reacceleration effects is the reason why the low energy spectra in our analysis are harder than those in Refs.~\cite{Donato:2008yx,Ibarra:2013qt}.
In Refs.~\cite{Donato:2008yx,Ibarra:2013qt}, the authors solved the propagation equation by using the semi-analytical method~\cite{Donato:2001ms} and the ``med'' parameter set following Ref.~\cite{Donato:2003xg}.
In those analyses, the convection and reacceleration effects are simultaneously taken into account.
A large Alfven velocity $52.9\mathrm{km}/\mathrm{s}$ would also lead to more antideuterons in the sub-GeV region than our predictions.

With the uncertainties from the hadronic model and propagation model, the predicted background antideuteron flux is still below the expected sensitivities of AMS-02 and GAPS.
Therefore, if the cosmic antideuterons are detected by these experiments in the future, there should be some exotic sources of antideuterons.

\subsection{DM Antideuterons}
The DM annihilations to hadronic final states can contribute to the antideuteron flux through the parton shower and hadronization processes.
We employ the generator PYTHIA 8~\cite{Sjostrand:2007gs} to simulate the DM annihilations to $b\bar{b}$ final states.
For the coalescence momentum, we adopt a typical $p_0$ value of $192\MeV$ following Ref.~\cite{Ibarra:2012cc}, which has been tuned for explaining the $\bar{d}$ measurement of ALEPH~\cite{Schael:2006fd}.

The CR contribution of DM annihilation depends on the DM density distribution in the Galaxy.
We adopt the widely used Navarro-Frenk-White (NFW)~\cite{Navarro:1996gj} profile
\begin{equation}
  \rho(r) = \frac{\rho_0}{\left( r/r_s \right)\left( 1+r/r_s \right)^2},
  \label{equation:NFW}
\end{equation}
with $\rho_0=0.35\GeV\cm^{-3}$ and $r_s=20\kpc$.
Considering the $95\%$ limits on the DM annihilation cross section $\langle\sigma v\rangle$ derived from the AMS-02 $\bar{p}/p$ observation~\cite{Lin:2016ezz}, we estimate the maximal allowed values of the DM antideuteron flux.
For a DM with a mass of $50\GeV$, $100\GeV$ and $1000\GeV$, such limits are $(0.93,3.10,8.51)\times10^{-26}\mathrm{cm^3s^{-1}}$, respectively.

\begin{figure}[!htp]
  \centering
  \includegraphics[height=0.5\textwidth]{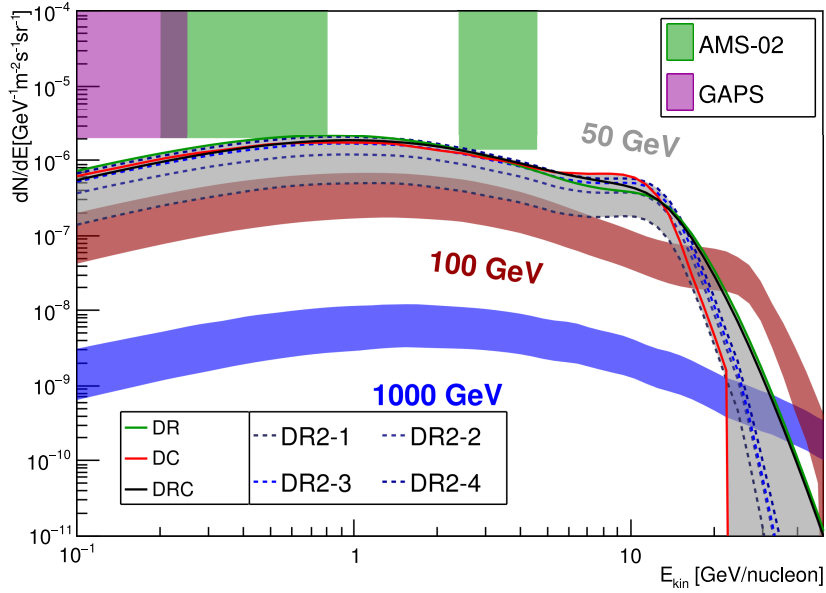}
\caption{The antideuteron flux from DM in $b\bar{b}$ channel.
All the lines are estimated from DM with mass $50\GeV$ for different propagation cases.
These lines would span an uncertainty band.
The gray, blue, red band are correspondingly the uncertainty band for DM mass of $50\GeV$, $100\GeV$ and $1000\GeV$.
The $\langle \sigma v\rangle$ used is the corresponding $95\%$ limit constrained by $\bar{p}/p$ data, adopted from \cite{Lin:2015taa}, $\langle \sigma v\rangle_{50,100,1000}=(0.93,3.10,8.51)\times10^{-26}\mathrm{cm^3s^{-1}}$.
}
  \label{fig:dbar_dm_prop}
\end{figure}

We show the results given by for different propagation models in Fig.~\ref{fig:dbar_dm_prop}.
We can see that the antideuteron spectra from DM are flatter than that of the background antideuteron in the sub-GeV region.
This is because the DM antideuterons in this energy region are dominantly generated in the boosted jets from DM annihilations; there is no significant low energy cut in the injection spectrum.
Thus the antideuterons losing energies via the convection and reacceleration effects would not significantly affect the final flux at low energies.
As the background antideuteron background is very small, it is promising to search for the DM signatures in the sub-GeV region.

We also find that unlike the background antideuteron, the propagation coefficient $D_0$ and the halo size $L$ would significantly affect the antideuteron flux from DM\@.
This difference is caused by the different distributions of the DM and background antideuteron sources.
The DM antideuterons are generated by DM annihilations in the entire DM halo, while the background antideuterons are produced by the collisions between the primary CRs and interstellar gas in the Galaxy disk.
Therefore, the propagation uncertainty of the DM antideuteron flux dominantly comes from the propagation halo size.
With these uncertainties considered, we find that a DM with a mass smaller than tens of $\GeV$ would result in a detectable antideuteron signal at future experiments.

\section{Conclusion}\label{section_conclusion}
In this work, we provide a sensible calculation on uncertainties of the secondary antideuteron flux.
These uncertainties estimation is necessary for the solid determination of the DM antideuteron signature in the future.

In the analysis, we pay special attention to the uncertainties from the hadronic interaction models, and use the MC generators EPOS and QGSJET-II to calculate the injection of background antideuterons.
The uncertainties from different propagation model are also carefully studied.
Considering constraints from the current B/C ratio data, several typical propagation cases are chosen in our analysis.

Taking the $p$ and $\bar{p}/p$ observation data into account, we finally show the uncertainties of the background antideuteron flux from the propagation and hadronic interaction models separately.
The background antideuteron fluxes predicted with different MC generators would vary by a factor of $\sim 2$ around several $\GeV$, while those predicted with different propagation models would vary by a factor of $\sim 3-10$ in the sub-GeV region.

In addition, we also discuss the propagation uncertainty for the antideuteron flux from DM annihilation\@.
Unlike background antideuterons, DM antideuterons are sensitively affected by the propagation halo size $L$.

With all the uncertainties involved, we compare the expectations with the prospected experiment sensitivities.
We find that the antideuteron background is still difficult to observe, while the annihilation signatures from DM particles with masses around tens of $\GeV$ are possible to detect at future detectors.

\begin{acknowledgements}
This work is supported by the National
Key Program for Research and Development (No. 2016YFA0400200) and by the National
Natural Science Foundation of China under Grants No. U1738209, 11475189, 11475191.
\end{acknowledgements}

\bibliography{anti_deuteron}
\end{document}